\documentclass[twocolumn,showpacs,preprintnumbers,amsmath,amssymb,epsf]{revtex4}

\usepackage{graphicx}
\usepackage{dcolumn}
\usepackage{bm}

\begin{document}


\title{Phase transition in a one-dimensional Ising ferromagnet at
    zero-temperature under Glauber dynamics with a synchronous updating
    mode
}

\author{Katarzyna Sznajd--Weron}
\email{kweron@ift.uni.wroc.pl} \homepage{http://www.ift.uni.wroc.pl/~kweron}
\affiliation{Institute of Theoretical Physics, University of Wroc{\l}aw, pl. Maxa
Borna 9, 50-204 Wroc{\l}aw, Poland }

\date{\today}

\begin{abstract}
In the past decade low-temperature Glauber dynamics for the one-dimensional Ising system has been several times observed experimentally and occurred to be one of the most important theoretical approaches in a field of molecular nanomagnets. On the other hand, it has been shown recently that Glauber dynamics with the Metropolis flipping probability for the zero-temperature Ising ferromagnet under synchronous updating can lead surprisingly to the antiferromagnetic steady state. In this paper the generalized class of Glauber dynamics at zero-temperature will be considered and the relaxation into the ground state, after a quench from high temperature, will be investigated. Using Monte Carlo simulations and a mean field approach, discontinuous phase transition between ferromagnetic and antiferromagnetic phases for a one-dimensional ferromagnet will be shown.   
\end{abstract}


\maketitle

\section{Introduction}
Glauber dynamics for the Ising spin chain has been known for almost 50 years \cite{Glauber}, but only recently it became a really hot topic, not only from a fundamental, but also an applicative point of view \cite{CGLSSVVRPN2001,CGLSSVVRPN2002,GS2004,CCLWM2004,BSSG2005,BBCGS2006,MJYC2009}. It is well known that a purely one-dimensional (1D) system exhibits long-range ordering only at zero temperature $T=0K$. Nevertheless, in some situations long relaxation times for the magnetization reversal with decreasing temperature can be observed, and finally at significantly low temperatures, the material can behave as a magnet. The phenomenon of slow magnetic relaxation is considered as one of the most important achievements of molecular magnetism, opening exciting new perspectives including that of storing information \cite{GCNS1993,GCPS1994}. Slow relaxation of the magnetization, predicted in the 1960s by Glauber in a chain of ferromagnetically coupled Ising spins \cite{Glauber}, in materials composed of magnetically isolated chains was observed for the first time in 2001 \cite{CGLSSVVRPN2001}. In 2002, this new class of nanomagnets was named single-chain magnets (SCM) \cite{CGLSSVVRPN2002} (for a recent review see \cite{MJYC2009}) and the Glauber dynamics for the one-dimensional Ising spins system became one of the most important theoretical approaches for SCM. 

Within the Glauber dynamics for Ising spins with a spin $s=1/2$, in a broad sense, each spin is flipped $S_i(t) \rightarrow -S_i(t+1)$ with a rate $W(\delta E)$ per unit time and this rate is assumed to depend only on the energy difference implied in the flip. In this paper we consider the generalize class of zero-temperature dynamics defined as:
\begin{eqnarray}
W(\delta E)=
\left\{
\begin{array}{ll}
1 & \mbox{if }  \delta E<0, \\ 
W_0 & \mbox{if }  \delta E=0, \\ 
0 & \mbox{if }  \delta E>0, 
\end{array}
\right.
\label{w}
\end{eqnarray}
which occurred to be very interesting not only from an applicative perspective, but also from a theoretical point of view as an example of non-equilibrium dynamical systems with many attractors \cite{GL04}. The zero-temperature limits of the original Glauber dynamics \cite{Glauber} and Metropolis rates \cite{metrop} (two the most popular choices) are respectively $W^{G}_0 = 1/2$ and $W^M_0 = 1$. 

Glauber dynamics was originally introduced as a sequential updating (SU) process \cite{Glauber}. Also Monte Carlo method, used frequently for various models in statistical physics, as proposed originally by Metropolis et al. \cite{metrop}, is essentially SU process. Evolution under dynamics defined by (\ref{w}) with random sequential updating is already well known in a case of one-dimensional system and can be derived analytically \cite{GL04}.  For any non-zero value of the rate $W_0$ ferromagnetic steady state is reached and  the dynamics belongs to the universality class of the zero-temperature Glauber model \cite{Glauber}. The particular value $W_0 = 0$ corresponds to the constrained zero-temperature Glauber dynamics (\cite{GL04} and references therein). In the constrained zero-temperature Glauber dynamics, the only possible moves are flips of isolated spins and therefore the system  eventually reaches a blocked configuration, where there is no isolated spin \cite{GL04}, i.e. for $W_0=0$ the relaxation time to the ferromagnetic steady state is infinite.

The case of the synchronous updating, in which all units of the system are updated at the same time, is much more interesting. Moreover, clear evidence of a relaxation mechanism which involves the simultaneous reversal of spins have been shown experimentally for magnetic chains at low temperatures \cite{BCFGMNPRSV2004}. 

\begin{figure}
\begin{center}
\includegraphics[scale=0.4]{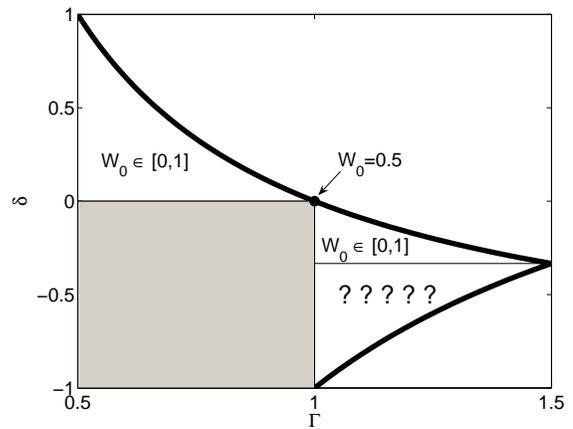}
\caption{Thick lines correspond to equations $\delta=(1-\Gamma)/\Gamma$ and $\delta=(\Gamma-2)/\Gamma$. The region between these two lines corresponds to the condition $W(\delta E) \in [0,1]$. In \cite{MO2000} the region denoted by the gray color has been investigated (i.e. $\delta<0,\Gamma \in (0,1)$), while in this paper we investigate two regions signatured by $W_0 \in [0,1]$. The region denoted by '?????' has not been investigated up till now, and it is not covered by the one-parameter model considered in this paper.}
\label{fig_pd}
\end{center}
\end{figure}

In \cite{MO2000} more general form of zero-temperature Glauber dynamics has been investigated than one defined by equation by (\ref{w}). They have studied a model with two parameters $\Gamma$ and $\delta$, which can be presented  at $T=0$ analogously to (\ref{w}) as:
\begin{eqnarray}
W(\delta E)=
\left\{
\begin{array}{ll}
\Gamma(1+\delta) & \mbox{if }  \delta E<0, \\ 
\frac{\Gamma}{2}(1-\delta) & \mbox{if }  \delta E=0, \\ 
0 & \mbox{if }  \delta E>0, 
\end{array}
\right.
\label{wg}
\end{eqnarray}
where again $W(\delta E)$ denotes the flipping rate per unit time. To fulfill the condition $W(\delta E) \in [0,1]$,
as seen from equation (\ref{wg}), the following relations have to be satisfied :
\begin{eqnarray}
-1 \le \delta \le \frac{1-\Gamma}{\Gamma} \nonumber\\
\frac{\Gamma-2}{\Gamma} \le \delta \le 1
\end{eqnarray}
Above relations correspond to the region between thick lines in Fig.\ref{fig_pd}. In \cite{MO2000} only the region denoted by the gray color in Fig.\ref{fig_pd} has been investigated (i.e. $\delta<0,\Gamma \in (0,1)$). Comparing equations (\ref{w}) and (\ref{wg}) we can easily derive the following relations:
\begin{eqnarray}
\Gamma & = & W_0 +\frac{1}{2} \nonumber\\
\delta & = & \frac{1/2-W_0}{1/2+W_0}.
\label{rel12}
\end{eqnarray}
In this paper we consider one-parameter model defined by Eq.(\ref{w}) with $W_0 \in [0,1]$. Using relations (\ref{rel12}) we can determine corresponding regions in Fig.\ref{fig_pd} (signatured by $W_0 \in [0,1]$), which are disjoint from the gray region investigated in \cite{MO2000}. Also the area denoted by '?????' in Fig.\ref{fig_pd} has not been investigated up till now -- it could be considered using two-parameter model defined by (\ref{wg}), but it is not covered by the one-parameter model, which is a subject of this paper. 

\section{Simulation and mean field results}
We consider the chain of $L$ Ising spins $\sigma_i = \pm 1 \; (i=1,2,\ldots L)$ with the periodic boundary conditions. In the initial state each lattice site is occupied independently by a randomly chosen value $+1$ or $-1$, both equally probable (high temperature situation). In every time step all spins are considered simultaneously, but each spin is flipped independently with probability $W(\delta E)$ defined by Eq. (\ref{w}).
It occurs that for all $W_0 \in (0,1)$ system eventually reaches one of the two final states - ferromagnetic steady state or antiferromagnetic limit cycle. If we measure the density of bonds (bond connects two sites with opposite spins):
\begin{equation}
\rho=\frac{1}{L} \sum_{i=1}^L (1-\sigma_i \sigma_{i+1}),
\label{ro}
\end{equation}
we obtain in the final state $\rho_{st}=1$ (antiferromagnetic state) or $\rho_{st}=0$ (ferromagnetic state).

\begin{figure}
\begin{center}
\includegraphics[scale=0.4]{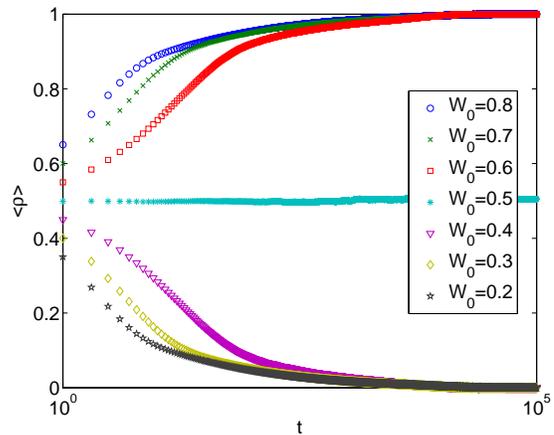}
\caption{The time evolution of the mean value of the density of bonds $<\rho>$ measured in Monte Carlo steps for the lattice size $L=160$ is presented. Averaging was done over $10^4$ samples. For $W_0<0.5$ the mean number of bonds decreases in time to $0$ (ferromagnetic steady state) and for $W_0>0.5$ increases to $1$ (antiferromagnetic limit cycle).}
\label{ab_time}
\end{center}
\end{figure}

The time evolution of the mean value (averaged over $10^4$ samples) of the density of bonds measured in Monte Carlo steps (MCS) is presented in Fig.\ref{ab_time}. This is seen that for $W_0<0.5$  the average number of bonds decreases in time and eventually the system reaches the ferromagnetic steady state ($<\rho(\infty)>=<\rho_{st}>=0$), while for $W_0>0.5$ it increases and eventually  antiferromagnetic limit cycle is reached ($<\rho(\infty)>=<\rho_{st}>=1$). Results presented in Fig. \ref{ab_time} show that for $W_0=0.5$ there is a phase transition between ferromagnetic and antiferromagnetic phase. 

This phase transition can be predicted using the mean field approximation analogously as it was done in \cite{MO2000}.
In \cite{MO2000} the mean field equations for the density of active bonds and magnetization has been derived:
\begin{eqnarray}
\frac{d\rho}{dt} & = & 2 \delta \Gamma \rho (1-3\rho+2\rho^2) \nonumber \\
\frac{dm}{dt} & = & -\delta\Gamma m(m^2-1). 
\end{eqnarray}
Using relations (\ref{rel12}) we can easily rewrite above equations in the case of our one-parameter model:
\begin{eqnarray}
\frac{d\rho}{dt} & = & (1-2W_0) \rho (1-3\rho+2\rho^2) \nonumber \\
\frac{dm}{dt} & = & (W_0-\frac{1}{2}) m(m^2-1). 
\label{mfa}
\end{eqnarray}
As we see there are three types of fixed points:
\begin{eqnarray}
m_{st}  =  \pm 1 & \mbox{and} & \rho_{st}  =  0 \nonumber \\
m_{st}  =  \pm 0 & \mbox{and} & \rho_{st}  =  1/2 \nonumber\\
m_{st}  =  \pm 0 & \mbox{and} & \rho_{st}  =  1 \nonumber
\end{eqnarray}
In \cite{MO2000} only two first types have been considered:
\begin{itemize} 
\item $\rho_{st}=0$ (ferromagnetic state with $m_{st}=-1,1$)
\item $\rho_{st}=1/2$ (so called active phase). 
\end{itemize}
However, there is a third fixed point $\rho_{st}=1,m_{st}=0$, which corresponds to antiferromagnetic steady state found in our computer simulations. It can be easily checked that:
\begin{itemize} 
\item 
for $W_0<0.5$ ferromagnetic fixed point ($m_{st}=\pm 1, \rho_{st}=0$) is stable
\item for $W_0>0.5$ antiferromagnetic fixed point ($m_{st}=0,\rho_{st}=1$) is the stable one.
\end{itemize}
Thus, a mean field approximation predicts discontinuous phase transition between ferromagnetic and antiferromagnetic phase for $W_0=0.5$. This should be noticed that the transition value $W_0=0.5$ corresponds to the original Glauber dynamics \cite{Glauber}. 

\begin{figure}
\begin{center}
\includegraphics[scale=0.4]{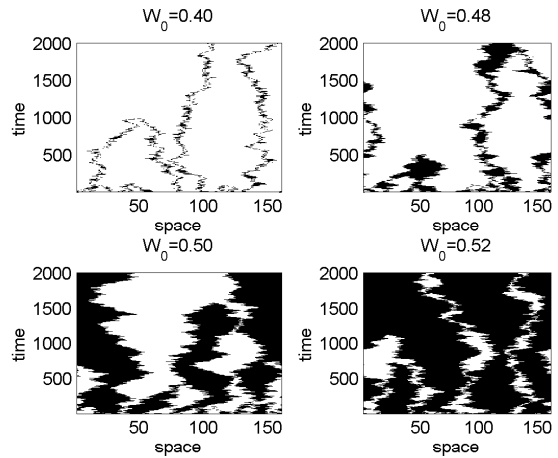}
\caption{The time evolution of the Ising spins chain of the length $L=160$ is presented. Black points represent bonds and thus black regions correspond to antiferromagnetic and white to ferromagnetic clusters. Coexistence of both types of clusters is visible for $W_0 \approx 0.5$. For $W_0=0.5$ both types of clusters are nearly the same size and there is a long-time competition between them leading eventually to one of two possible steady states (ferromagnetic or antiferromagnetic)}
\label{map}
\end{center}
\end{figure}

In the case of discontinuous phase transition one would expect the phase coexistence. We have provided computer simulations to confirm this mean field result and indeed coexistence of ferro- and antiferromagnetic phases can be observed near the transition point $W_0=0.5$ (see Fig.\ref{map}). For $W_0=0.5$ both types of clusters (ferro- and antiferromagnetic) are nearly the same size and after a long-time competition between them eventually one of two possible steady states is reached. Because for $W_0=0.5$ both of them are equally probable we see the constant value of the average density of bonds in Fig.\ref{ab_time}. Let us now investigate the phase transition more quantitatively using Monte Carlo Simulations. 

\begin{figure}
\begin{center}
\includegraphics[scale=0.4]{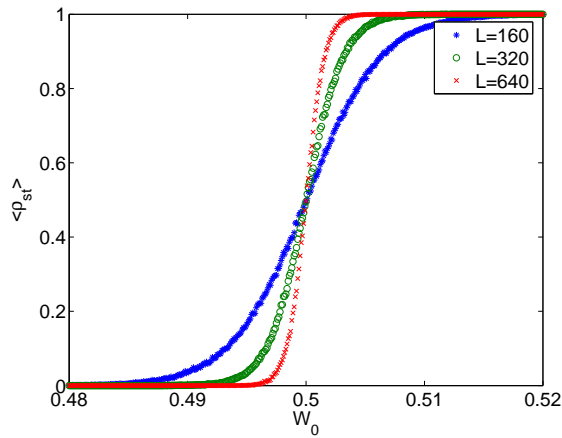}
\caption{Density of bonds $\rho_{st}$ in stationary state as a function of flipping probability $W_0$ (so called exit probability) averaged over $10^4$ samples. In the thermodynamical limit $L \rightarrow \infty$ for $W_0<0.5$ ferromagnetic steady state is reached with probability one ($\rho_{st}=0$) and $W_0<0.5$ antiferromagnetic steady state is reached with probability one ($\rho_{st}=1$). Note that, the transition value $W_0=0.5$ corresponds to the original Glauber dynamics.}
\label{ro_w}
\end{center}
\end{figure}

Following \cite{MO2000,RVM2007} we use as an order parameter the mean value of the density of bonds. We provide Monte Carlo simulations and wait until the system reaches the final stationary state. Dependence between order parameter in the stationary state $<\rho_{st}>$ and the flipping probability $W_0$ is presented in Fig. \ref{ro_w}, showing again clearly discontinuous phase transition for $W_0=0.5$ in agreement with the mean field result. In the case of $W_0<0.5$ the ferromagnetic steady state is obtained with probability $1$ (for the infinite system $L=\infty$ ). For $W_0>0.5$ the antiferromagnetic state is always reached, i.e. the stationary states losses any remnants of the ferromagnetic Ising interactions.   

\begin{figure}
\begin{center}
\includegraphics[scale=0.4]{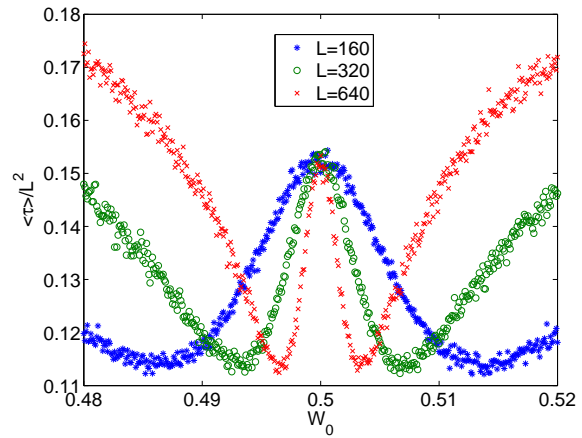}
\caption{The mean relaxation times $<\tau>$ divided by the square of lattice size $L$ as a function of flipping probability $W_0 \in [0.48,0.52]$. Averaging was done over $10^4$ samples. Note that for $W_0=0.5$ relaxation time scales with the system size as $<\tau> \sim L^2$. However, for $W_0 \ne 0.5$ scaling exponent differs from known value $\alpha=2$ (see Fig.6).}
\label{tau2}
\end{center}
\end{figure}

One of the most important issues connected with the coarsening is the relaxation time $\tau$, i.e. time needed to reach the ground state. In this paper we measure the relaxation time starting from the random initial conditions and counting how many Monte Carlo steps is needed to reach the steady state ($\rho=1$ or $\rho=0$). We average over $N=10^4$ samples and calculate the mean relaxation time:
\begin{equation}
<\tau>=\frac{1}{N} \sum_{i=1}^N \tau_i,
\end{equation}
where $\tau_i$ is the relaxation time of $i$-th sample.
In Fig.\ref{tau2} $<\tau>$ divided by the square of the lattice size $L$ as a function of the flipping probability $W_0$ is shown. This is seen that for $W_0=0.5$ the mean relaxation time scales as $\tau \sim L^2$, which is well known result in a case of sequential updating \cite{B1994,RK1998}. The dependence between the mean relaxation time $<\tau>$ and the flipping probability $W_0$ is non-monotonical. For $W_0 \rightarrow 0$ the relaxation time grows rapidly, which can be understood recalling that $<\tau>$ if infinite for $W_0=0$ \cite{GL04}. For increasing $W_0$ the mean relaxation time decreases up to a certain point $W_0^{min}(L)$. However, due to the phase transition in $W_0=0.5$, for $W_0 \in (W_0^{min}(L),0.5)$ it grows again, resulting non-monotonic behavior shown in Fig.\ref{tau2}. The maximum peak is more and more narrow with the growing lattice size, which is expected behavior for the phase transition. The minimal value $W_0^{min}(L)$ depends on the system size $L$ as $W_0^{min}(L)=-2.5/L + 0.5$ and therefore $\lim_{L \rightarrow \infty} W_0^{min}(L) \rightarrow 0.5$. The mean relaxation time for this minimal value scales with the system size as $<\tau(W_0^{min})> \sim L^2$, i.e. with the same exponent as for the transition point $W_0=0.5$. 

The most important question here is the one concerning the origin of the phase transition. 
As it was mentioned above, in the case of Metropolis flipping rate ($W_0=1$) the system reaches antiferromagnetic limit cycle, instead for the ferromagnetic steady state \cite{BB2004,RVM2007}. It can be easily understood, because for the flipping probability $W_0=1$, the case of synchronous updating is fully deterministic (see an example below):
\begin{eqnarray}
\cdots \uparrow \uparrow \uparrow \downarrow \downarrow \downarrow \cdots \nonumber\\
\cdots \uparrow \uparrow \downarrow \uparrow \downarrow \downarrow \cdots \nonumber\\
\cdots \uparrow \downarrow \uparrow \downarrow \uparrow \downarrow \cdots \nonumber\\
\cdots \downarrow \uparrow  \downarrow \uparrow  \downarrow \uparrow  \cdots \nonumber\\
\cdots \uparrow \downarrow \uparrow \downarrow \uparrow \downarrow \cdots 
\end{eqnarray}

On the other hand, only for $W_0=1$ updating is really synchronous. For decreasing $W_0$ only isolated spins are concerned really synchronously, since in the case of isolated spins $\delta E<0$ (see equation (\ref{w})) the flip is provided with the probability $1$. Flipping of isolated spins leads clearly to growth of ferromagnetic domains. 
Let us introduce for a while a notation $L_{\delta E=0}$ for the number of spins that flipping would not change the energy and $L_{\delta E<0}$ for the number of spins that flipping would decrease the energy. The flip for $\delta E=0$ is realized with the probability $W_0$ and for $\delta E<0$ with the probability $1$, which means that on average $L_{\delta E<0}+W_0L_{\delta E=0}$ is flipped in a single time step. In the case of $W_0=1$, as mentioned above, the antiferromagnetic order is reached. On the other hand, for $W_0=1/L_{\delta E=0}$ on average only one not isolated spin (i.e. with $\delta E=0$) is flipped in a single time step, similarly to the case of the sequential updating for the system without isolated spins. Thus, because in the case of sequential updating ferromagnetic steady state is reached, one can expect also ferromagnetic order in the case of synchronous updating for small values of $W_0$. Clearly the phase transition must occur somewhere between the antiferromagnetic order, preferred by a fully synchronous updating ($W_0=1$), and the ferromagnetic steady state, preferred by sequential updating ($W_0=1/L_{\delta E=0}$).   
 
\begin{figure}
\begin{center}
\includegraphics[scale=0.4]{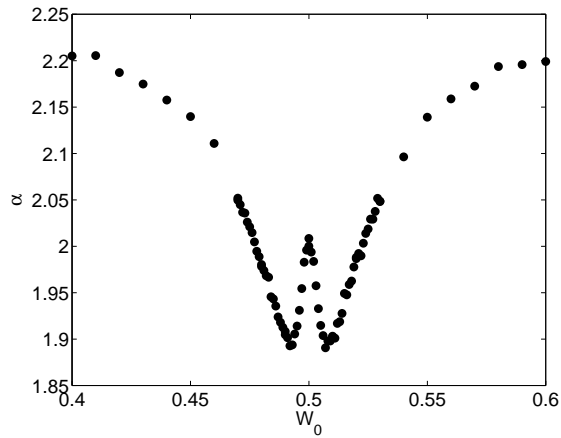}
\caption{The mean relaxation time $<\tau>$ scales with the system size as $<\tau> \sim L^{\alpha}$. For $W_0=0.5$ the scaling exponent $\alpha=2$, which is well known result in the case of sequential updating. However, in general scaling exponent depends on the flipping probability $W_0$, i.e. $\alpha=\alpha(W_0)$. Dependence between scaling exponent $\alpha$ and the flipping probability $W_0$ is shown. Simulations were done for the system size $L \in [20,1280]$ and averaged over $10^4$ samples}.
\label{scal}
\end{center}
\end{figure}

As mentioned above, for $W_0=0.5$ and $W_0=W_0^{min}(L)$ the mean relaxation time scales with a system size as $\sim L^2$. We have checked also the scaling for other values of $W_0$ and we have obtained power laws  $<\tau(W_0)> \sim L^{\alpha}$ with $W_0$-depending scaling exponents $\alpha=\alpha(W_0)$. The dependence between scaling exponent and the flipping probability is presented in Fig. \ref{scal}. The shape of the curve $\alpha(W_0)$ mimic the shape of $<\tau(W_0)>$, which can be understood looking at Fig. \ref{tau2}. 

\section{Summary}
In this paper we have been investigating the relaxation of the Ising spins chain under the generalized class of Glauber dynamics at zero-temperature. Within such a dynamics, the flipping probability in a case of conserved energy is given by arbitrary value of $W_0 \in [0,1]$ (review in a case of sequential updating can be find in \cite{GL04}). We have proposed to use synchronous updating for such a generalized class of zero-temperature dynamics. Our motivation for this work came from recent experiments showing slow relaxation in magnetic chains at low temperatures \cite{CGLSSVVRPN2001,CGLSSVVRPN2002,GS2004,CCLWM2004,BSSG2005,BBCGS2006,MJYC2009,BCFGMNPRSV2004}.
We have shown by Monte Carlo simulations that there is a phase transition for $W_0=0.5$, which correspond to the value originally proposed by Glauber \cite{Glauber}. Following \cite{MO2000} we were able to obtain the mean field result which predicts discontinuous transition between ferro- and antiferromagnetic phases for $W_0=0.5$. 

\acknowledgments 
I would like to thank prof. Geza Odor for fruitful discussions and paying my attention to the very interesting paper \cite{MO2000}.


\begin{thebibliography}{33}
\bibitem{Glauber}
R.J. Glauber, J. Math. Phys. {\bf 4}, 294 (1963)
\bibitem{CGLSSVVRPN2001}
A. Caneschi, D. Gatteschi, N. Lalioti, C. Sangregorio, R. Sessoli, G. Venturi, A. Vindigni, A. Rettori, M. G. Pini and M. A. Novak, Angew. Chem., Int. Ed. Engl. {\bf 40}, 1760 (2001)
\bibitem{CGLSSVVRPN2002}
A. Caneschi, D. Gatteschi, N. Lalioti, C. Sangregorio, R. Sessoli, G. Venturi, A. Vindigni, A. Rettori, M. G. Pini and M. A. Novak, Europhys. Lett. {\bf 58}, 771 (2002)
\bibitem{GS2004}
D. Gatteschi and R. Sessoli, Journal of Magnetism and Magnetic Materials 272–276 (2004) 1030–1036
\bibitem{CCLWM2004}
C. Coulon, R. Cl\'erac, L. Lecren, W. Wernsdorfer and H. Miyasaka, Phys. Rev.B {\bf 69}, 132408 (2004)
\bibitem{BSSG2005}
L. Bogani, C. Sangregorio, R. Sessoli, and D. Gatteschi, Angew.
Chem., Int. Ed. Engl. {\bf 44}, 5817 (2005)
\bibitem{BBCGS2006}
K. Bernot, L. Bogani, A. Caneschi, D. Gatteschi, and R. Sessoli,
J. Am. Chem. Soc. {\bf 128}, 7947 (2006)
\bibitem{MJYC2009}
H. Miyasaka, M. Julve, M. Yamashita and R. Cl\'erac, Inorg. Chem. {\bf 48} 3420 (2009)
\bibitem{GCNS1993}
D. Gatteschi, A. Caneschi, M. A. Novak and R. Sessoli, Nature {\bf 365} 141 (1993)
\bibitem{GCPS1994}
D. Gatteschi, A. Caneschi, L. Pardi and R. Sessoli, Science {\bf 265} 1054 (1994)
\bibitem{GL04}
C. Godreche, J. M. Luck, Journal of Physics - Condensed Matter, {\bf 17}, 2573 (2005)
\bibitem{metrop}
N. Metropolis, A.W. Rosenbluth, M.N. Rosenbluth, A.H.Teller, E. Teller, J. Chem. Phys. 21 (1953) 1087.
\bibitem{BB2004}
D. Boll\'e, B. Busquets, Eur. Phys. J. B {\bf 42}, 397 (2004)
\bibitem{RVM2007}
F. Radicchi, D. Vilone and H. Meyer-Ortmanns, J Stat Phys {\bf 129}, 593 (2007)
\bibitem{BCFGMNPRSV2004}
L. Bogani, A. Caneschi, M. Fedi, D. Gatteschi, M. Massi, M. A. Novak, M. G. Pini,
A. Rettori, R. Sessoli and A.Vindigni, Phys. Rev. Lett. {\bf 92}, 207204 (2004)
\bibitem{B1994} 
A.J. Bray, Adv. Phys. {\bf 43}, 357 (1994)
\bibitem{RK1998}
S. Redner and P. L. Krapivsky, J. Phys. A 31, 9929 (1998) 
\bibitem{VRJ2004}
A. Vindigni, N. Regnault, and Th. Jolicoeur, Phys. Rev. B {\bf 70}, 134423 (2004)
\bibitem{PR2007}
M. G. Pini and A. Rettori, B {\bf 76}, 064407 (2007)
\bibitem{MO2000}
N. Menyhard and G. Odor, Brazilian J. of Physics, {\bf 30}, 113 (2000) 
\end{thebibliography}
\end{document}